\documentclass[aps,two column]{revtex4}
\usepackage{eurosym}
\usepackage{amsfonts}
\usepackage{amsmath}
\usepackage{amssymb,epsf}
\usepackage{color}
\usepackage{hyperref}
\usepackage{orcidlink}

\begin{document}

\title{$k-$deformed Schwarzschild AdS black strings}
\author{B. Eslam Panah\,\orcidlink{0000-0002-1447-3760}$^1$,$^2$}
\email{eslampanah@umz.ac.ir}
\affiliation{$^1$ Department of Theoretical Physics, Faculty of Basic Sciences, University of Mazandaran, P. O. Box 47416-95447, Babolsar, Iran.\\
$^2$ Center for Theoretical Physics, Khazar University, 41 Mehseti Str., Baku, AZ1096, Azerbaijan.}

\begin{abstract}
Considering a cylindrically symmetric spacetime, a novel exact $\kappa$-deformed Schwarzschild-AdS black string solution is derived in General Relativity in the presence of a negative cosmological constant. The geometric properties of the resulting solution are systematically investigated by computing the Ricci and Kretschmann curvature scalars. The analysis reveals the existence of a curvature singularity at the origin, which remains enclosed by an event horizon provided that the $\kappa$-deformation parameter satisfies a specific constraint. Subsequently, the influence of various model parameters on the horizon structure and configuration is examined in detail. The thermodynamic properties of these black strings are thoroughly evaluated; in this regard, the Hawking temperature, entropy, mass, isobaric heat capacity, and Gibbs free energy are self-consistently extracted. The effects of the parameters, particularly the $\kappa$-deformation parameter, on these thermodynamic quantities are comprehensively explored. Furthermore, it is demonstrated that these state variables fundamentally satisfy the extended first law of black hole thermodynamics and yield a modified Smarr relation. Finally, the temporal profile of the Hawking radiation is probed via the dimensionless sparsity parameter, through which the impacts of the underlying parameters on the quantum emission rate are systematically analyzed.

\end{abstract}

\maketitle

\section{\textbf{Introduction}}

Asymptotically anti-de Sitter (AdS) spacetimes occupy a central position in
contemporary gravitational physics, largely because of their fundamental
role in the AdS/CFT correspondence and their effectiveness as a precise
theoretical arena for examining the intricate interplay among spacetime
geometry, horizon structure, and thermodynamic behavior. In this setting,
exact solutions of Einstein's field equations with nontrivial horizon
topologies--including topological black holes, cylindrical geometries, and
other extended gravitational configurations--have been studied extensively,
revealing features that differ substantially from those of the standard
spherically symmetric case \cite{Surya2001,Cardoso2001,Morley2018}. As a
result, such geometries provide valuable frameworks for testing matter
couplings, modified gravitational dynamics, and semiclassical effects \cite%
{Ali2020,Deglmann2025,Ahmed2025,Barbosa2025,Santos2026}.

Among these configurations, black strings constitute a particularly
important class of extended gravitational objects. First constructed by
Lemos in four-dimensional AdS spacetime \cite{Lemos1995}, black strings
describe cylindrically symmetric solutions in which the event horizon
extends continuously along a longitudinal direction. In contrast to ordinary
black holes, whose horizons are compact, black strings possess noncompact
horizon topology and therefore exhibit geometric and thermodynamic
properties that are qualitatively distinct from those of their spherical
counterparts. This extended structure makes them especially suitable for
investigating how horizon topology and spacetime asymptotics influence
gravitational dynamics.

The study of cylindrical collapse further enhances the importance of black
string geometries, since such systems provide a natural setting for
analyzing prolate collapse scenarios and their associated physical
implications \cite{Thorne1972}. Over the years, black strings have been
investigated in a wide range of contexts, including classical gravitational
dynamics, thermodynamic phase structure, perturbative stability, geodesic
motion, and analogue gravity models \cite%
{Sa1996,Boonserm2019,Ghosh2020,Hendi2021,Jusufi2023,Ma2024,Lessa2025,Darlla2025,Alencar2026a,Chen2026,Rodriguez2026,Alencar2026b}. These
developments continue to demonstrate that black strings offer a productive
framework for understanding the role of matter content, asymptotic
structure, and horizon extension in strong-gravity systems \cite%
{Kumar2023,Pereira2025}.

While General Relativity successfully describes a wide range of gravitational phenomena, it is widely believed to fail in extreme, strong-field regimes. Under such conditions, quantum fluctuations become prominent, highlighting the need for a consistent quantum gravity framework to fully describe physics of compact objetcs such as black holes and black strings. Numerous models have been proposed to bridge the gap between classical gravitation and quantum mechanics, aiming to uncover the fundamental nature of spacetime.  In this context, non-commutative (NC) spacetime geometry \cite{Maresca2025} provides a prominent framework for encoding quantum gravitational corrections. In this setup, the spacetime fabric is characterized by an underlying NC algebra, which inherently sets a minimal fundamental length scale \cite{Snyder1947,Doplicher1994}. Furthermore, recent developments suggest that such NC structures emerge naturally from perturbative quantum gravity near the Planck scale \cite{Frob2023a,Frob2023b}. Over the past decades, various NC spacetime models have been explored to incorporate these quantum features into gravity, with the Moyal spacetime \cite{Douglas2001} and $\kappa$-deformed spacetime \cite{Arzano2021} serving as the two most representative examples.

One of interesting NC backgrounds is related to the  $\kappa$-deformed spacetime which characterized by a Lie-algebraic structure. Defined by the commutation relations
$[\hat{x}^i, \hat{x}^0] = i a \hat{x}^i, \quad [\hat{x}^i, \hat{x}^j] = 0$, where $a$ denotes the $\kappa$-deformation parameter, this framework emerges intuitively within theories of Doubly Special Relativity (DSR) \cite{Kowalski-Glikman2002}. The fundamental symmetry of these models is encapsulated by the $\kappa$-Poincare algebra \cite{Lukierski1992}, which has also been identified as a feature of the infrared limit of loop quantum gravity (LQG) \cite{Cianfrani2016}. This algebraic structure inherently modifies the standard Heisenberg algebra and introduces deviations from conventional dispersion relations \cite{Lukierski1995}. Interestingly, research has extended these concepts to black hole physics, successfully describing NC generalizations of BTZ and Kerr black holes using $\kappa$-deformed algebras \cite{Dolan2007,Schupp2009}. Formulating gravity in this setting is notably more complex than in Moyal spacetime, primarily due to the intricate nature of the underlying Lie-algebraic geometry \cite{Rozental2025}. Despite these difficulties, several investigations have successfully modeled $\kappa$-deformed black hole solutions. A commonly employed methodology, the realization approach, maps non-commutative functions into commutative variables alongside their conjugate momenta and the deformation parameter \cite{Gupta2014,Harikumar2017}. While earlier work has yielded significant insights (such as evaluating $\kappa$-deformed corrections to Bekenstein-Hawking entropy and Hawking radiation \cite{Gupta2014,Harikumar2017,Gupta2015,Juric2016,Gupta2023}) a comprehensive thermodynamic framework for these objects is currently lacking. Specifically, detailed investigations into phase transitions, criticality, and the systematic derivation of the first law of thermodynamics within an extended phase space remain open problems for future study.

In this work, we derive the $\kappa$-deformed Schwarzschild-AdS black string solution within the framework of Einstein gravity. To this end, we employ a regularization scheme that replaces the localized point-like singularity with a smeared mass distribution, thereby incorporating NC corrections into the energy-momentum tensor \cite{Nicolini2006, Kobakhidze2009}. Specifically, by solving the Poisson equation with the $\kappa$-modified Newtonian gravitational potential \cite{Rajagopal2025},
\begin{equation}
V(r) = -\frac{M}{r} \left( 1 - \frac{a M}{12\pi r^2} \right), \label{eq:potential}
\end{equation}
where $a$ denotes the $\kappa$-deformation parameter in natural units, we determine the effective energy density of the spacetime. Consequently, the non-vanishing components of the energy-momentum tensor are obtained as
\begin{equation}
T_{0}^{~0} = T_{1}^{~1} = -\frac{aM^{2}}{8\pi ^{2}r^{5}}, \quad T_{2}^{~2} = T_{3}^{~3} = \frac{3aM^{2}}{16\pi ^{2}r^{5}}, \label{DyTensor}
\end{equation}
which clearly exhibit the inherent anisotropy of the effective fluid, with $T_{0}^{~0} = -\rho$, $T_{1}^{~1} = P_{r}$, and $T_{2}^{~2} = T_{3}^{~3} = P_{\varphi} = P_{z}$ denoting the energy density, radial pressure, and transverse pressures, respectively. 

Notably, the anisotropic fluid induced by this $\kappa$-deformation satisfies the null, weak, and strong energy conditions, while violating the dominant energy condition \cite{Kumara2026}. Such a behavior is fully consistent with black hole configurations modified by LQG, thereby reinforcing the physical viability of our solution within well-established quantum gravity frameworks \cite{Muniz2025}.

\section{\textbf{Field Equation and Black String Solutions}}

The field equations of Einstein-$\Lambda$ gravity are given by
\begin{equation}
G_{\mu }^{~\nu }+\Lambda \delta _{\mu }^{~\nu }=8\pi T_{\mu }^{~\nu }, \label{FE}
\end{equation}
where $G_{\mu }^{~\nu }=\mathcal{R}_{\mu }^{~\nu }-\frac{1}{2}\delta _{\mu }^{~\nu }\mathcal{R}$ denotes the Einstein tensor, with $\mathcal{R}_{\mu }^{~\nu }$ and $\mathcal{R}$ representing the Ricci tensor and the Ricci scalar curvature, respectively. Throughout this study, we work in natural units with $c=G=1$, where $c$ is the speed of light in vacuum and $G$ is the Newtonian gravitational constant. The cosmological constant, denoted by $\Lambda$, governs the asymptotic structure of the spacetime, yielding a dS background for $\Lambda > 0$ or an AdS geometry for $\Lambda < 0$.

To explore the properties of black strings, we consider a four-dimensional, static, and cylindrically symmetric spacetime. Following the pioneering construction by Lemos \cite{Lemos1995}, the metric preserving time-reversal and translational symmetries along the longitudinal axis is parametrized as
\begin{equation}
ds^{2}=-f(r)dt^{2}+\frac{dr^{2}}{f(r)}+r^{2}\left(d\varphi ^{2}+\alpha ^{2}dz^{2}\right), \label{metric}
\end{equation}
where $f(r)$ represents the metric potential to be determined from the field equations. Here, $\alpha$ is a constant parameter with dimensions of inverse length ($[\alpha]=L^{-1}$), associated with the characteristic AdS curvature radius.

Substituting the energy-momentum tensor components from Eq. \eqref{DyTensor}
and the metric ansatz \eqref{metric} into the gravitational field equations %
\eqref{FE}, the explicit components of the equations of motion are derived
as follows 
\begin{eqnarray}
eq_{tt} &=&eq_{rr}=\Lambda r^{2}+rf^{\prime }\left( r\right) +f\left(
r\right)  \notag \\
&&+\frac{aM^{2}}{\pi r^{3}},  \label{eqtt} \\
&&  \notag \\
eq_{\varphi \varphi } &=&eq_{zz}=r^{2}f^{\prime \prime }\left( r\right)
+2rf^{\prime }\left( r\right) +2\Lambda r^{2}  \notag \\
&&-\frac{3aM^{2}}{\pi r^{3}},  \label{eqthethe}
\end{eqnarray}
where the prime and double prime denote the first and second derivatives with respect to $r$, respectively. Also, in the above expressions, the terms $eq_{tt}$, $eq_{rr}$, $eq_{\varphi \varphi}$, and $eq_{zz}$ represent the respective non-vanishing components of the gravitational field equations \eqref{FE} associated with the $t$, $r$, $\varphi$, and $z$ coordinates.

By decoupling the field equations \eqref{eqtt} and \eqref{eqthethe}, the functional form of the metric coefficient $f(r)$ is analytically determined as 
\begin{equation}
f\left( r\right) =\frac{-2M}{r}+\frac{aM^{2}}{2\pi r^{3}}-\frac{\Lambda r^{2}%
}{3}.  \label{F(r)}
\end{equation}

\subsection{\textbf{Curvature Scalars}}

To explore the geometric properties of the resulting spacetime, we compute the Ricci and Kretschmann curvature invariants, which provide essential insights into its curvature structure and singular behavior.

The explicit analytical expression for the Ricci scalar ($\mathcal{R}$) is obtained as

\begin{equation}
\mathcal{R}=4\Lambda -\frac{aM^{2}}{\pi r^{5}},  \label{R}
\end{equation}%
An asymptotic analysis of the Ricci scalar reveals that
\begin{eqnarray}
\underset{r\rightarrow 0}{\lim }\mathcal{R} &\rightarrow &\infty ,
\label{R1} \\
&&  \notag \\
\underset{r\rightarrow \infty }{\lim }\mathcal{R} &\rightarrow &4\Lambda ,
\label{R2}
\end{eqnarray}%
As indicated by Eq.~\eqref{R1}, the Ricci scalar diverges at the origin ($r=0$), signifying the presence of a physical curvature singularity. Furthermore, Eq.~\eqref{R2} demonstrates that the asymptotic structure of this geometry is dominated exclusively by the cosmological constant, thereby recovering the expected AdS curvature background at spatial infinity.

Furthermore, the Kretschmann scalar ($K \equiv \mathcal{R}_{\mu \nu \rho \sigma} \mathcal{R}^{\mu \nu \rho \sigma}$), which provides an essential quadratic measure of the spacetime curvature, is derived as
\begin{equation}
K(r) = \frac{8\Lambda ^{2}}{3} - \frac{4\Lambda a M^{2}}{3\pi r^{5}} + \frac{48M^{2}}{r^{6}} - \frac{80a M^{3}}{\pi r^{8}} + \frac{46a^{2}M^{4}}{\pi ^{2}r^{10}}. \label{K}
\end{equation}
An examination of the central and asymptotic limits of the Kretschmann invariant yields 
\begin{align}
\lim_{r\rightarrow 0}K& \rightarrow \infty ,  \label{K1} \\
\lim_{r\rightarrow \infty }K& =\frac{8\Lambda ^{2}}{3}.  \label{K2}
\end{align}%
The divergence at $r=0$ demonstrated in Eq.~\eqref{K1} confirms the existence of a true physical curvature singularity at the core of the spacetime, which cannot be eliminated by coordinate transformations. Moreover, in the far-field regime ($r \to \infty$), the result in Eq.~\eqref{K2} shows that the geometry asymptotically settles into a maximally symmetric AdS background.

\subsection{\textbf{Visualizing the Metric Function and Event Horizon}}

Before investigation the location of the event horizons which is identified
through the zeros of the metric coefficient $f(r)$, we can find an upper
limit on the $\kappa -$deformation parameter $a$. For this purpose, we
re-write the metric function (\ref{F(r)}), in the following form 
\begin{equation}
P\left( r\right) =r^{5}+\frac{6M}{\Lambda }r^{2}-\frac{3aM^{2}}{2\pi \Lambda 
},  \label{P(r)}
\end{equation}%
where $P\left( r\right) $ is obtained by multiple $-\frac{3r^{2}}{\Lambda }$
in the metric function, i.e., $P(r)=-\frac{3r^{2}}{\Lambda }f(r)$.

The behavior of $P\left( r\right) $ indicate that:

\begin{itemize}
\item[i)] \textbf{Near the origin} ($r\rightarrow 0$): The dominate term is
the positive term $-\frac{3aM^{2}}{2\pi \Lambda }$, because $\Lambda <0$. In
other words, the value of the metric function is always positive due to the
existence of the AdS background.

\item[ii)] A\textbf{s $r$ increases}: The negative term $\frac{6M}{\Lambda }%
r^{2}$ (arising from the gravitational attraction effect of the mass $M$)
causes the function to decrease and descend, until it reaches its minimum
value (the bottom of the valley) at the point denoted by $r_{\text{min}}$.

\item[iii)] \textbf{With further increase of $r$}: The term $r^{5}$ (arising
from the AdS background curvature) becomes dominant, causing the function to
rise once again and tend toward $+\infty $.
\end{itemize}

Now, we should determine the location $r_{\text{min}}$ of the function $%
P\left( r\right) $ (or the metric function). For this purpose, we consider $%
P^{\prime }\left( r\right) =0$, which leads to $r_{\text{min}}=\left( \frac{%
-12M}{5\Lambda }\right) ^{1/3}$. By replacing $r=r_{\text{min}}$ within Eq. (%
\ref{P(r)}), and by imposing the condition $P\left( r_{\text{min}}\right)
\leq 0$, we find an upper limit on $\kappa -$deformation parameter $a$ in
the following form 
\begin{equation}
a\leq \frac{12\pi }{5M^{1/3}}\left( \frac{-12}{5\Lambda }\right) ^{2/3},
\end{equation}%
where is is positive for AdS background. Also, the value of $\kappa -$%
deformation parameter $a$ increases by decreasing $M$.

It is notable that, the condition $P\left( r_{\text{min}}\right) >0$,
reveals that there is a naked singularity for the metric function (\ref{F(r)}%
). There is an exterm root for the metric fucntion (\ref{F(r)}), when $%
P\left( r_{\text{min}}\right) =0$. In addition, there exist two roots (an
event horizon and a Cauchy horizon) when $P\left( r_{\text{min}}\right) <0$.

\begin{figure}[tbph]
\centering
\includegraphics[width=0.7\linewidth]{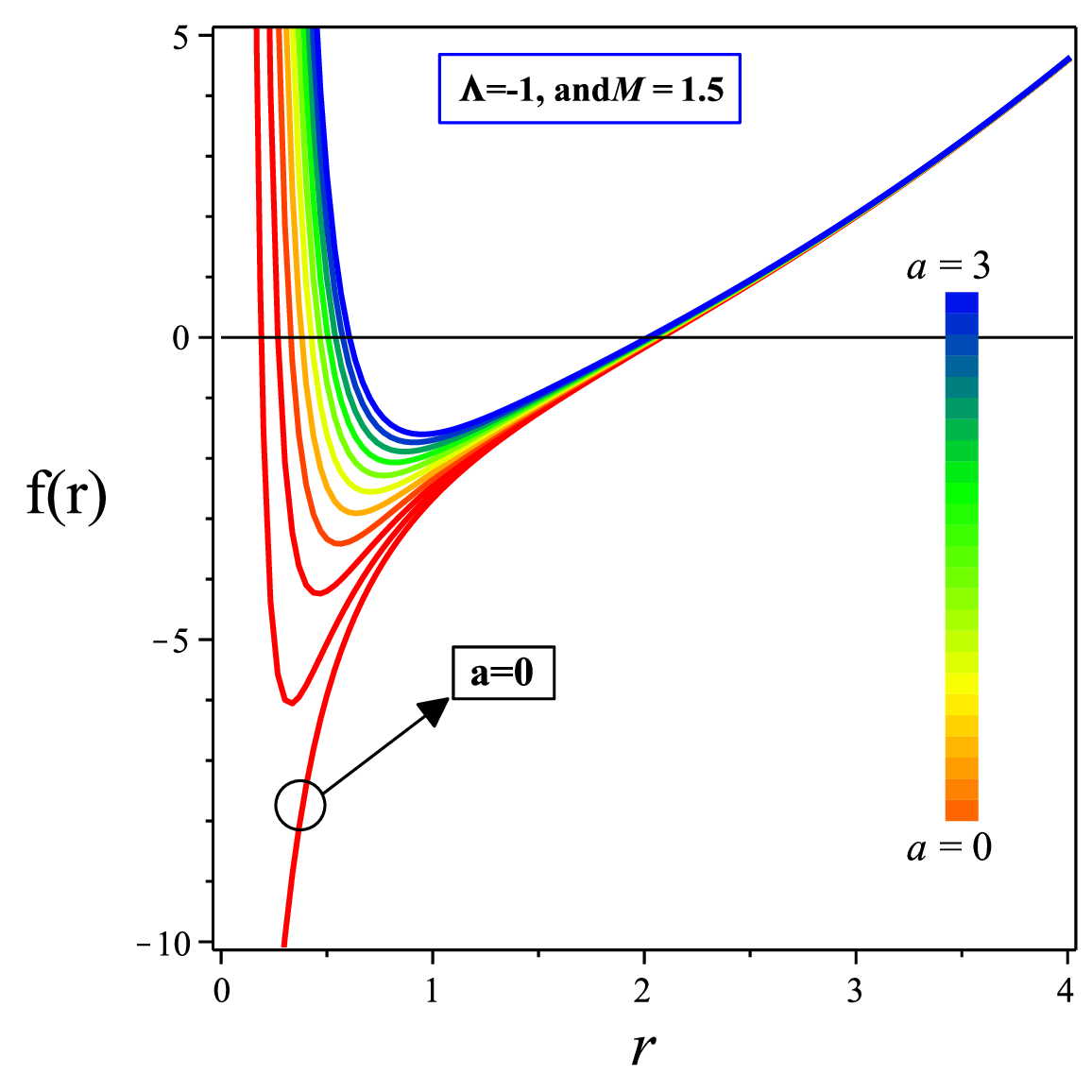} \includegraphics[width=0.7%
\linewidth]{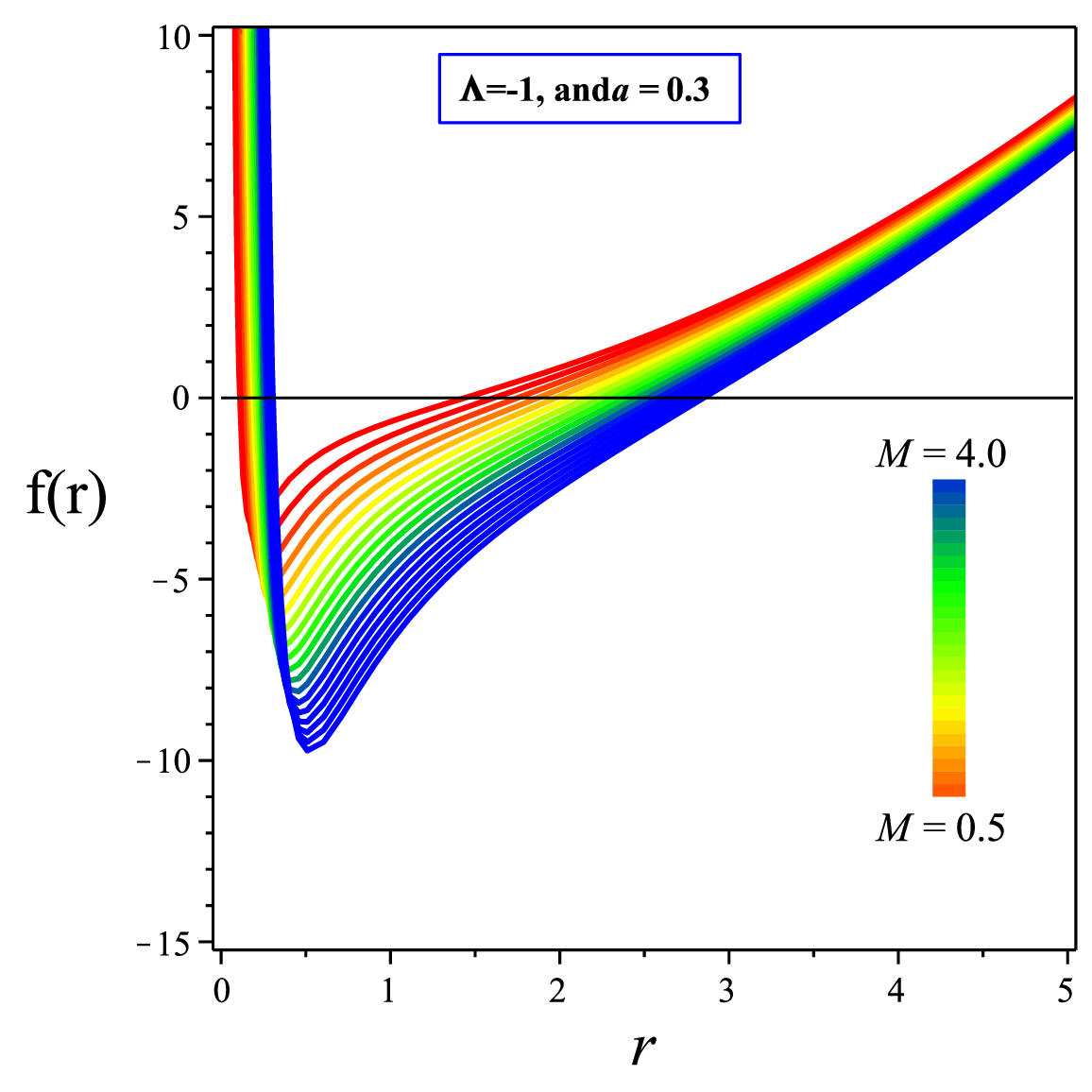}
\caption{The metric function $f(r)$ plotted as a function of $r$ for varying
values of $a$ (up panel), and $M$ (down panel).}
\label{Fig1}
\end{figure}

We study the effects of $\kappa -$deformation parameter $a$ and mass $M$ on
the roots of the metric function in Fig. \ref{Fig1}. Our findings reveal
that:

\begin{itemize}
\item[i)] \textbf{Th effect of $\kappa-$deformation parameter $a$}: there is
one root and it is related to the even horizon when $a=0$ (denoted by $%
r_{+_{a=0}}$). In other words, in the absence of $\kappa-$deformation
parameter $a$, there is one root. By applying the $\kappa -$deformation
parameter $a$, two roots appear. The small and large roots are related to
the Cauchy and the event horzions, respectively. By increasing $a$, the
small and large roots decreases (see the up panel in Fig. \ref{Fig1}). In
other words, the large event horizon belongs to black strings without $%
\kappa-$deformed case.

\item[ii)] \textbf{The effect mass $M$}: Large event horizon belongs to
massive black strings. In other words, by increasing the mass ($M$), the
event horizon increases (see the down panel in Fig. \ref{Fig1}).
\end{itemize}

\section{\textbf{Thermodynamic}}

The present section is dedicated to a comprehensive characterization of the thermodynamic profile of $\kappa$-deformed Schwarzschild-AdS black strings. Our investigation centers on the derivation of fundamental quantities, specifically the Hawking temperature and entropy, while concurrently assessing the system's local and global thermal stabilities. Furthermore, we explore the sensitivity of these thermodynamic variables to adjustments in the model's parameters, thereby demarcating the physical domains characterized by stable and unstable phases. Finally, we demonstrate that these quantities satisfy the first law of thermodynamics and, consequently, derive the corresponding Smarr relation.

The Wald entropy of these black strings is given by 
\begin{equation}
\widetilde{S}=\frac{\Omega r_{+}^{2}}{4}\left( 1+\frac{4ar_{+}P}{9}\right) ,
\label{S0}
\end{equation}%
where 
\begin{equation}
\Omega =\left. \frac{1}{r^{2}}\int_{0}^{2\pi }\int_{0}^{l_{z}}\sqrt{%
g_{\varphi \varphi }g_{zz}}\,d\varphi \,dz\right\vert _{r=r_{+}}=2\pi \alpha
l_{z}.  \label{Omega}
\end{equation}%
To obtain a quantity independent of the longitudinal scale, we define the
entropy per unit effective string length ($\alpha l_{z}$). By combining Eqs.
(\ref{S0}) and (\ref{Omega}), the reduced entropy density $S$ is derived as 
\begin{equation}
S=\frac{\widetilde{S}}{\alpha l_{z}}=\frac{\pi r_{+}^{2}}{2}\left( 1+\frac{%
4ar_{+}P}{9}\right) ,  \label{S}
\end{equation}%
where $S$ deremtines by the pressure $P$, and $\kappa -$deformation
parameter $a$. The entropy of these black strings is always positive when $%
a>0$.

By supposing the $\kappa -$deformed Schwarzschild AdS black strings modifed
the event horizon up to first order of $a$, in the following form 
\begin{equation}
r_{+}=r_{0}\left( 1+\delta a\right) ,  \label{r+}
\end{equation}%
where $r_{0}=\sqrt{\frac{4S}{\Omega }}$ is the event horizon in the standard
form of Schwarzschild black strings when $a=0$. By replacing Eq. (\ref{r+})
within the Wald entropy (Eq. (\ref{S})), we find that 
\begin{eqnarray}
S &=&\frac{\Omega r_{0}^{2}\left( 1+\delta a\right) ^{2}}{4}+\frac{\Omega
ar_{0}^{3}P\left( 1+\delta a\right) ^{3}}{9}  \notag \\
&&  \notag \\
&=&\frac{\Omega r_{0}^{2}\left( 1+2\delta a\right) }{4}+\frac{\Omega
ar_{0}^{3}P}{9}+\mathcal{O}\left( \delta \left( a^{2}\right) \right) ,
\end{eqnarray}%
by neglecting higher-order terms $\mathcal{O}\left( a^{2}\right) $, and
applying $\left( 1+\delta a\right) ^{2}\simeq 1+2\delta a$, we get 
\begin{eqnarray}
S &=&\frac{\Omega r_{0}^{2}}{4}+\frac{\Omega r_{0}^{2}\delta a}{2}+\frac{%
\Omega ar_{0}^{3}P}{9}  \notag \\[1pt]
&&  \notag \\
&=&\frac{\Omega r_{0}^{2}}{4}+a\left( \frac{\Omega r_{0}^{2}\delta }{2}+%
\frac{\Omega r_{0}^{3}P}{9}\right) ,
\end{eqnarray}%
by ignoring the term $a$, i.e., $\frac{\Omega r_{0}^{2}\delta }{2}+\frac{%
\Omega r_{0}^{3}P}{9}=0$, we can obtain 
\begin{equation}
\delta =-\frac{2r_{0}P}{9},  \label{delta}
\end{equation}%
where indicate that by considering Eq. (\ref{delta}) within Eq. (\ref{r+}),
the Wald entropy is equal with Bekenstein-Hawking area law.

Considering Eqs. (\ref{delta}) and (\ref{r+}), we can re-write the event
horizon versus the pressure ($P$), and the $\kappa -$deformation parameter $%
a $ in the following form 
\begin{equation}
r_{+}=r_{0}\left( 1-\frac{2ar_{0}P}{9}\right) .  \label{r+final}
\end{equation}

The ADM mass $\widetilde{\mathcal{M}}$ of the $\kappa -$deformed AdS black
string is obtained by 
\begin{equation}
\widetilde{\mathcal{M}}=\frac{M}{2}.  \label{M0}
\end{equation}%
where $M$\ is getting by solving the metric function ($f(r_{+})=0$) which
leads to 
\begin{equation}
M=\frac{2\Omega r_{+}^{3}P}{3}\left( 1+\frac{ar_{+}P}{3}\right) ,  \label{M1}
\end{equation}%
by replacing Eq. (\ref{M1}) within Eq. (\ref{M0}), we can get the total mass
of these black strings per unit effective string length ($\alpha l_{z}$) in
the following from 
\begin{equation}
\mathcal{M}\left( r_{+},P,a\right) =\frac{\widetilde{\mathcal{M}}}{\alpha
l_{z}}=\frac{2\pi r_{+}^{3}P}{3}\left( 1+\frac{ar_{+}P}{3}\right) .
\label{M}
\end{equation}

Similar to the entropy, the total mass depends on the pressure $P$, and $%
\kappa -$deformation parameter $a$. By increasing $a$ and $P$, the total
mass increases. In addition, the total mass is always positive when $a>0$.
At the origin ($r_{+}\rightarrow 0$), the total mass reaches to zero, i.e., $%
\underset{r_{+}\rightarrow 0}{\lim \mathcal{M}}\rightarrow 0$.

Using Eq. (\ref{r+final}), we can re-write $r_{+}^{3}=r_{0}^{3}\left( 1-%
\frac{2ar_{0}P}{9}\right) ^{3}\simeq r_{0}^{3}\left( 1-\frac{2ar_{0}P}{3}%
\right) =r_{0}^{3}-\frac{2ar_{0}^{4}P}{3}$, and $r_{+}^{4}\simeq r_{0}^{4}$.
Then, by replacing them within the total mass (Eq. (\ref{M})), we find 
\begin{equation}
\mathcal{M}\left( r_{0},P,a\right) =\frac{2\pi r_{0}^{3}P}{3}-\frac{2\pi
ar_{0}^{4}P^{2}}{9},  \label{Mr0}
\end{equation}%
where the above relation is obtained by neglecting higher-order terms $%
\mathcal{O}\left( a^{i}\right) $ when $i\geq 2$.

Next, the temperature of the $\kappa$-deformed AdS black strings can be determined by applying the definition
\begin{eqnarray}
T &=&\left( \frac{\partial \mathcal{M}\left( r_{0},P,a\right) }{\partial S}%
\right) _{P,a}  \notag \\
&=&\left( \frac{\partial r_{0}}{\partial S}\right) _{P,a}\left( \frac{%
\partial \mathcal{M}\left( r_{0},P,a\right) }{\partial r_{0}}\right) _{P,a},
\end{eqnarray}%
substituting the horizon radius $r_{0}=\sqrt{\frac{2S}{\pi }}$ along with Eq. (\ref{Mr0}) yields 
\begin{equation}
T=2r_{0}P\left( 1-\frac{4ar_{0}P}{9}\right) .  \label{Tr0}
\end{equation}

On the other hand, the Hawking temperature can alternatively be derived from the standard surface gravity formalism 
\begin{equation}
\kappa =\sqrt{-\frac{1}{2}(\nabla _{\mu }\chi _{\nu })(\nabla ^{\mu }\chi
^{\nu })}=\frac{1}{2}f^{\prime }(r)\big|_{r=r_{+}},
\end{equation}%
where $\chi _{\mu }$ is the Killing vector field in the form $\chi _{\mu
}=(1,0,0,0)$. Consequently, the Hawking temperature $T_{H}$ is obtained
through the relation 
\begin{equation}
T_{H}=\frac{\kappa }{2\pi }=\frac{1}{4\pi }f^{\prime }(r)\big|_{r=r_{+}}.
\label{Temp1}
\end{equation}%
By substituting the metric function (\ref{F(r)}) and imposing the horizon
condition from Eq. (\ref{M1}), the Hawking temperature for the $\kappa -$%
deformed AdS black string is derived as 
\begin{equation}
T_{H}=2r_{+}P\left( 1-\frac{2ar_{+}P}{9}\right) .  \label{Temp2}
\end{equation}

Here, we want to indicate that the obtained temperature in Eqs. (\ref{Tr0})
and (\ref{Temp2}) are the same. For this purpose, we should re-write $r_{0}$
versus the event horizon. By using Eq. (\ref{r+final}), we find 
\begin{equation}
r_{0}=r_{+}\left( 1+\frac{2ar_{+}P}{9}\right) ,  \label{r0r+}
\end{equation}%
where we neglect the higher-order terms $\mathcal{O}\left( a^{i}\right) $
when $i\geq 2$. Now, by replacing $r_{0}$ form Eq. (\ref{r0r+}) within Eq. (%
\ref{Tr0}), we have 
\begin{eqnarray}
T &=&2r_{+}P\left( 1+\frac{2ar_{+}P}{9}\right) \left( 1-\frac{4ar_{+}P}{9}%
\right)  \notag \\
&&  \notag \\
&=&2r_{+}P\left( 1-\frac{2ar_{+}P}{9}\right) ,
\end{eqnarray}%
where indicates that $T=T_{H}$. Notably, in order to extract the above
relation, we neglect the higher-order terms $\mathcal{O}\left( a^{i}\right) $
when $i\geq 2$.

Our analysis indicate that there is one real root for the Hawking
temperature in the following form 
\begin{equation}
r_{+}=\frac{9}{2aP},
\end{equation}%
in which $T$ is positive before this root. In other words, the Hawking
temperature of these black strings is positive in the range $0<r_{+}<\frac{9%
}{2aP}$, and it is negative when $r_{+}>$ $\frac{9}{2aP}$. Study of the
entropy (Eq. (\ref{S})), the total mass (Eq. (\ref{M})), and the Hawking
temperature (Eq. (\ref{Temp2})), simoultaneously, reveals that these black
strings can be physical system when the event horizon is located in the
range $0<r_{+}<\frac{9}{2aP}$ due to the existence of the positive value of
the temperature.

The thermodynamical volum per unit effective string length ($\alpha l_{z}$)
is given by 
\begin{equation}
V=\left( \frac{\partial \mathcal{M}\left( r_{0},P,a\right) }{\partial P}%
\right) _{S,a}=\frac{2\pi r_{0}^{3}}{3}-\frac{4\pi ar_{0}^{4}P}{9},
\end{equation}%
by replacing $r_{0}$ (from Eq. (\ref{r0r+})), within the above relation, we
find that 
\begin{equation}
V=\frac{2\pi r_{+}^{3}}{3}.  \label{V}
\end{equation}

The conjugate potential associated with the $\kappa -$deformation parameter $%
a$ per unit effective string length ($\alpha l_{z}$) \ is given by 
\begin{equation}
\Phi _{a}=\left( \frac{\partial \mathcal{M}\left( r_{0},P,a\right) }{%
\partial a}\right) _{S,P}=-\frac{2\pi r_{0}^{4}P}{9}=-\frac{2\pi r_{+}^{4}P}{%
9},  \label{Phi}
\end{equation}%
where we consider $r_{+}^{4}=r_{0}^{4}$ due to neglect the higher-order
terms $\mathcal{O}\left( a^{i}\right) $ when $i\geq 2$.

Using the termodynamics quantities such as Eqs. (\ref{S}), (\ref{M}), (\ref%
{Temp2}), (\ref{V}), and (\ref{Phi}), the first law of thermodynamics is
valid as 
\begin{equation}
d\mathcal{M}\left( r_{+},P,a\right) =TdS+VdP+\Phi _{a}da.
\end{equation}

Our findings indicate that the thermodynamics quantities can satisfy the
standard first law of thermodynamics for $\kappa -$deformed AdS black string
without needs to modification factor which is reported in Ref. \cite
{Kumara2026}.

\subsection{\textbf{Smarr Relation}}

To derive the associated Smarr relation, it is necessary
to determine the scaling dimensions of all quantities entering the mass function.

For the $k-$deformed Schwarzschild AdS black strings under consideration in four-dimensional, static, and cylindrically symmetric spacetime, the length dimensions of the relevant quantities are $\left[ 
\mathcal{M}\right] =L$, $\left[ S\right] =L^{2}$, $\left[ P\right] =L^{-2}$,
and $\left[ \alpha \right] =L$, where $L$ denotes a characteristic length
scale. Under a uniform rescaling of the length scale, $L\longrightarrow
\lambda L$, the thermodynamic variables and model parameters transform
according to $\mathcal{M}\rightarrow \lambda \mathcal{M}$, $S\rightarrow
\lambda ^{2}S$, $P\rightarrow \lambda ^{-2}P$, and $\alpha \rightarrow
\lambda \alpha $.

Consequently, the mass function obeys the weighted-homogeneity relation 
\begin{equation}
\mathcal{M}\left( \lambda ^{2}S,\lambda ^{-2}P,\lambda \alpha \right) =\lambda \mathcal{M}\left(
S,P,\alpha \right),  \label{lambdaM}
\end{equation}%
which establishes that $\mathcal{M}$ is a weighted homogeneous function of degree one.

Applying Euler's theorem for homogeneous functions to Eq.~(\ref{lambdaM}) yields

\begin{equation}
\mathcal{M}={}2S\left( \frac{\partial \mathcal{M}}{\partial S}\right)
_{P,a}-2P\left( \frac{\partial \mathcal{M}}{\partial P}\right)
_{S,a}+a\left( \frac{\partial \mathcal{M}}{\partial a}\right) _{S,P}.
\label{EulerMass}
\end{equation}

By substituting the respective thermodynamic conjugate quantities into Eq. (\ref{EulerMass}), one arrives at the modified Smarr relation
\begin{equation}
\mathcal{M}=2ST-2PV+a\Phi _{a}.  \label{ModifiedSmarr}
\end{equation}

\subsection{\textbf{Local Stability}}

To assess the local thermodynamic stability of the system, we evaluate the heat capacity at constant pressure, which is defined as
\begin{equation}
C_{p}=T\left( \frac{dS}{dT}\right) _{P,a}=T\frac{\left( \frac{dS}{dr_{+}}%
\right) _{P,a}}{\left( \frac{dT}{dr_{+}}\right) _{P,a}}.
\end{equation}%
By utilizing the expressions for the entropy (Eq. (\ref{S})), and
the Hawking temperature (Eq. (\ref{Temp2})), the heat capacity at
constant pressure for the $\kappa -$deformed AdS black strings is
analytically obtained as 
\begin{equation}
C_{p}=\pi r_{+}^{2}\left( 1+\frac{8ar_{+}P}{9}\right) ,  \label{C}
\end{equation}%
where terms of higher order $\mathcal{O}(a^2)$ have been neglected.

Equation~(\ref{C}) demonstrates that the isobaric heat capacity depends explicitly on both the $\kappa$-deformation parameter $a$ and the thermodynamic pressure $P$, growing monotonically as $a$ or $P$ increases. In the framework of black hole and black string thermodynamics, local thermodynamic stability is guaranteed only when the Hawking temperature and the heat capacity are simultaneously positive ($T > 0$ and $C_{P} > 0$). Given that $C_{P}$ remains strictly positive for $a > 0$, the physical domain of local stability is entirely dictated by the positivity of the Hawking temperature. As established earlier, the temperature remains positive within the horizon domain $0 < r_{+} < \frac{9}{2aP}$. Consequently, this interval defines the locally stable thermodynamic region, whose physical extent broadens as either the deformation parameter $a$ or the pressure $P$ decreases.

\begin{figure}[tbph]
\centering
\includegraphics[width=0.7\linewidth]{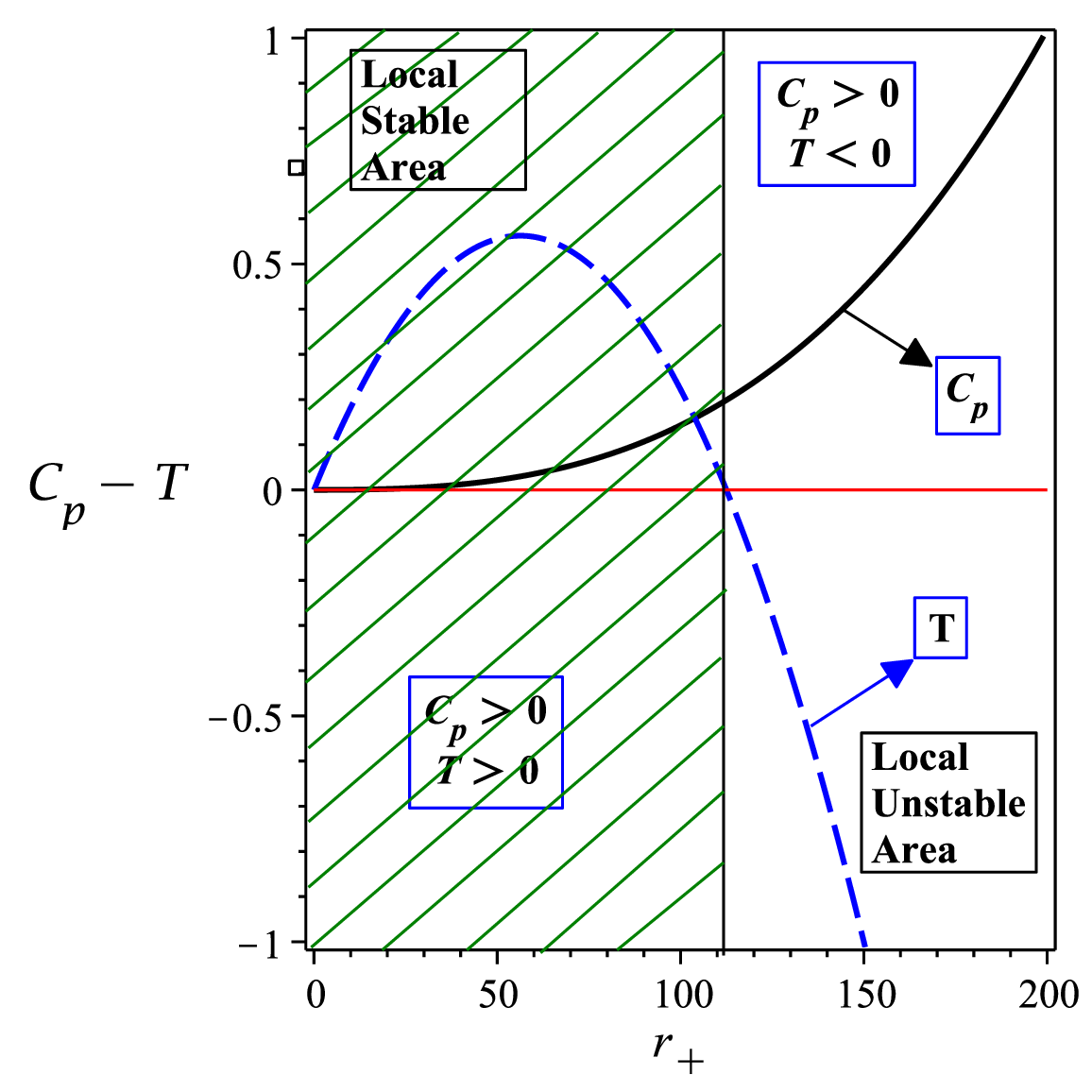}
\caption{The heat capacity $C_{p}$ (black continuous line) and the Hawking
temperature (blue dashed line) plotted as a function of $r_{+}$ for $%
\Lambda=-1$, $a=0.4$, and $P=0.1$.}
\label{Fig2}
\end{figure}

For illustrative purposes, the behavior of $C_{P}$ and $T$ as functions of the event horizon radius is displayed in Fig.~\ref{Fig2}. A vertical threshold line clearly delineates the stable and unstable thermodynamic regimes. The domain to the left of this boundary represents the physically stable phase, where both $C_{P} > 0$ and $T > 0$ are satisfied. Conversely, the region to the right of this boundary corresponds to a thermodynamically unstable phase characterized by an unphysical negative Hawking temperature.

\subsection{Global Stability}

In the context of the grand canonical ensemble, the global thermodynamic stability of a black hole or black string configuration is dictated by the behavior of its Gibbs free energy, where a negative value indicates thermodynamic preference over the reference background. Therefore, to determine the global stability domain, we evaluate the Gibbs free energy within the extended phase space, defined as
\begin{equation}
G = \mathcal{M} - TS.
\end{equation}

By substituting the explicit expressions for the entropy (Eq. (\ref{S})), the total mass (Eq.~(\ref{M})), and the Hawking temperature (Eq. (\ref{Temp2})), the Gibbs free energy of the $\kappa$-deformed AdS black string is obtained as
\begin{equation}
G = -\frac{\pi r_{+}^{3} P}{3}. \label{F}
\end{equation}
Remarkably, Eq. (\ref{F}) demonstrates that $G$ is strictly negative for all physical values of the horizon radius ($r_{+} > 0$) and pressure ($P > 0$). Furthermore, since $G(r_{+})$ possesses no non-trivial roots and remains non-vanishing throughout the physical domain, no Hawking-Page phase transition occurs. Consequently, the $\kappa$-deformed AdS black string configuration is globally thermodynamically stable everywhere within its physically admissible domain.

\subsection{Sparsity of Hawking Radiation}

Hawking radiation provides a vital probe of the quantum nature of black hole and black string physics. While conventionally characterized by the Hawking temperature and its associated spectral energy flux \cite{Page1,Page2}, recent investigations have highlighted that the evaporation process is intrinsically discrete rather than strictly continuous. In fact, Hawking emission proceeds as a sparse sequence of individual quanta whose temporal distribution is substantially modulated by the spacetime geometry and surrounding background fields \cite{Gray2016,Chowdhury2020,Silva2026}.

To assess the temporal characteristics of the emission from $\kappa$-deformed AdS black strings, we employ the dimensionless sparsity parameter introduced in Ref. \cite{Gray2016} 
\begin{equation}
\eta = \frac{\mathcal{C}}{\widetilde{g}} \frac{\lambda_{t}^{2}}{\widetilde{A}_{\mathrm{eff}}}, \label{eta}
\end{equation}
where $\widetilde{g}$ denotes the spin-degeneracy factor of the emitted quanta ($\widetilde{g}=1$ for scalar particles and $\widetilde{g}=2$ for massless bosons of non-zero spin \cite{Gray2016}), $\mathcal{C}$ is a dimensionless numerical coefficient, $\lambda_{t} = 2\pi/T$ is the characteristic thermal wavelength, and $\widetilde{A}_{\mathrm{eff}}$ represents the effective radiating area (geometric cross-section).

Following Page's optical geometric approach \cite{Page1}, the effective radiating area of a black hole is typically related to the horizon area via $\widetilde{A}_{\mathrm{eff}} = \frac{27}{4}\widetilde{A}_{\mathrm{BH}}$. Extending this geometric consideration to the cylindrical topology of black strings, the total effective area can be cast as $\widetilde{A}_{\mathrm{eff}} = \frac{27}{4}\widetilde{A}_{\mathrm{BS}}$, where $\widetilde{A}_{\mathrm{BS}} = 2\pi \alpha l_{z} r_{+}^{2}$ denotes the total horizon area. Consequently, the effective cross-sectional area per unit string length ($\alpha l_{z}$) is given by
\begin{equation}
A_{\mathrm{eff}} = \frac{\widetilde{A}_{\mathrm{eff}}}{\alpha l_{z}} = \frac{27\pi r_{+}^{2}}{2}. \label{Aeff}
\end{equation}

By taking $\mathcal{C}/\widetilde{g}$ in the standard normalization and substituting the thermal wavelength $\lambda_{t} = 2\pi/T$ along with Eq.~(\ref{Aeff}) into Eq.~(\ref{eta}), the sparsity parameter reduces to
\begin{equation}
\eta = \frac{8\pi}{27 r_{+}^{2} T^{2}}.
\end{equation}
Inserting the explicit expression for the Hawking temperature from Eq.~(\ref{Temp2}) and expanding up to first order in the deformation parameter, we obtain
\begin{equation}
\eta = \frac{2\pi}{27 r_{+}^{4} P^{2}} \left( 1 + \frac{4 a r_{+} P}{9} \right) + \mathcal{O}(a^2).
\end{equation}
Evidently, the sparsity of the radiated field is intimately governed by both the $\kappa$-deformation parameter $a$ and the thermodynamic pressure $P$. A detailed physical analysis reveals the following features:
\begin{itemize}
	\item[\textbf{(i)}] \textbf{Effect of the $\kappa$-deformation parameter $a$:} 
	The presence of the linear correction term $\left(1 + \frac{4ar_{+}P}{9}\right)$ indicates that for $a > 0$, the dimensionless sparsity parameter $\eta$ increases relative to the standard undeformed scenario. Physically, this implies that the average time delay between successively emitted wavepackets grows, rendering the quantum radiation process markedly sparser.
	
	\item[\textbf{(ii)}] \textbf{Effect of the thermodynamic pressure $P$:} 
	As the effective thermodynamic pressure $P$ increases, the sparsity parameter decreases monotonically as $\eta \propto P^{-2}$ to leading order. Consequently, in the high-pressure (or strong negative cosmological constant) regime, $\eta \to 0$, indicating that the wavepackets overlap significantly and the radiation process approaches a continuous, quasi-blackbody thermal beam.
\end{itemize}

\begin{figure}[tbph]
\centering
\includegraphics[width=0.7\linewidth]{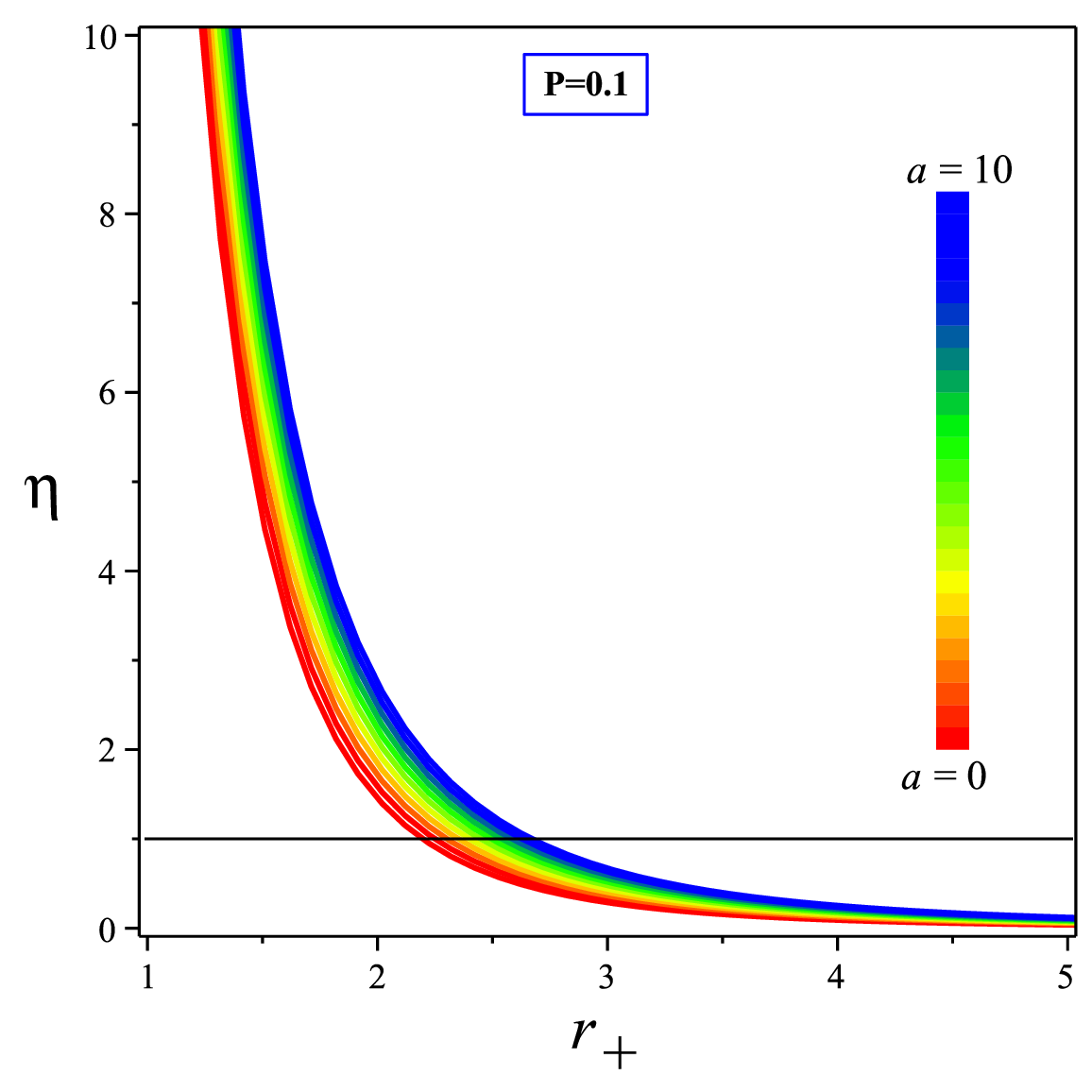} \includegraphics[width=0.7\linewidth]{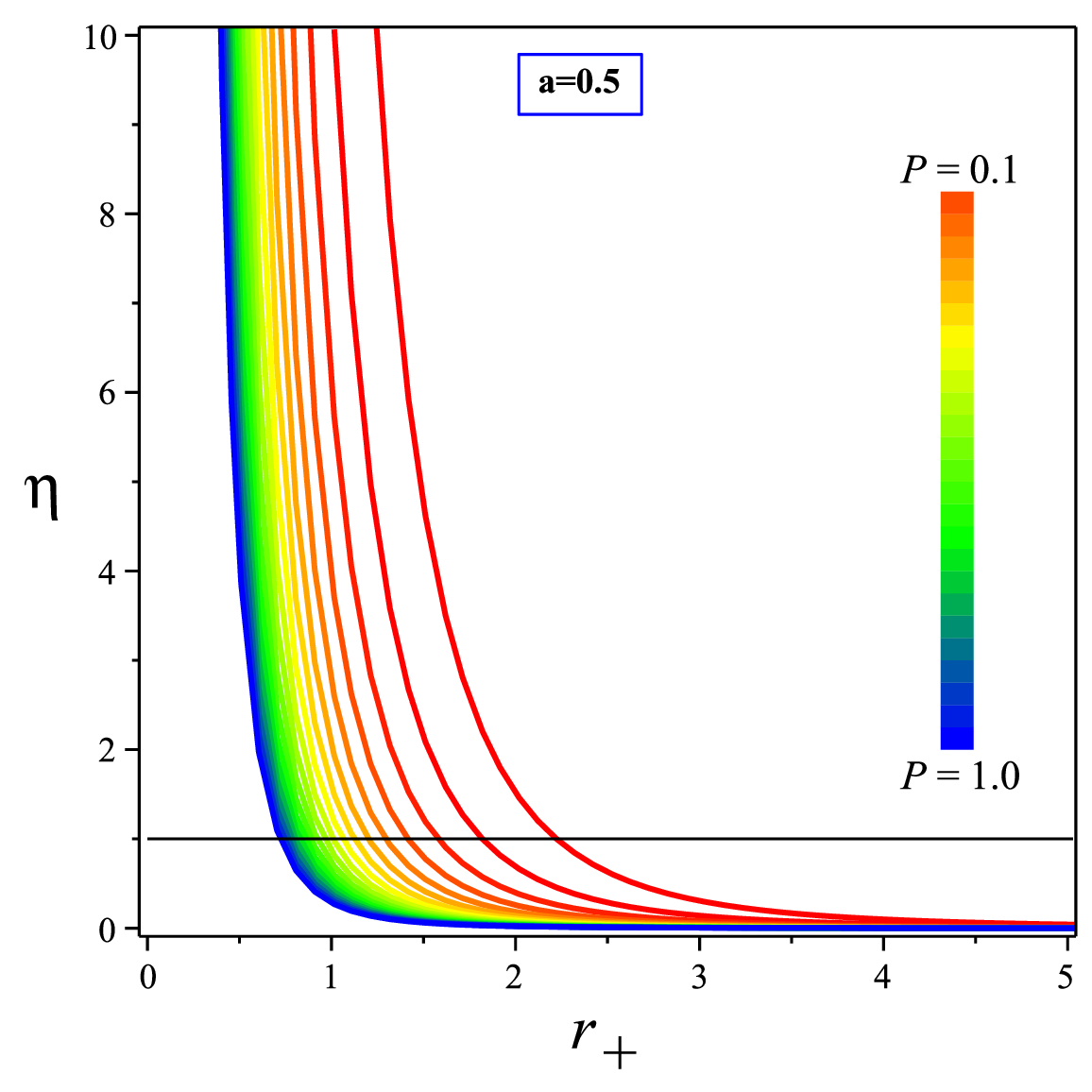}
\caption{The sparsity $\eta$ plotted as a function of $r_{+}$ for varying values of $a$ (up panel), and $P$ (down panel).}
\label{Fig3}
\end{figure}

\section{\textbf{Conclusion}}

In this work, the $\kappa$-deformed Schwarzschild-AdS black string solution has been derived within the framework of Einstein gravity. This has been achieved by employing a regularization scheme in which the localized point-like singularity is replaced with a smeared mass distribution, thereby incorporating NC corrections into the effective energy-momentum tensor. In particular, the energy density has been evaluated by solving the Poisson equation with the $\kappa$-modified Newtonian gravitational potential.

The geometric properties of the resulting spacetime have been investigated through the evaluation of both Ricci and Kretschmann curvature invariants. The divergence of these scalars at the origin ($r=0$) has confirmed the presence of an unavoidable, true physical curvature singularity at the core of the configuration. Furthermore, their asymptotic analysis has verified that the far-field geometry is exclusively dominated by the cosmological constant, thereby recovering the maximally symmetric AdS background at spatial infinity.

The structural properties and horizon configurations of $\kappa$-deformed AdS black strings have been systematically examined by constructing an auxiliary polynomial, wherein the existence of distinct horizons, extremal states, or naked singularities has been classified based on the deformation parameter $a$.

It has been revealed that the introduction of the NC parameter splits the standard commutative horizon into two concentric horizons, with their radii observed to shrink as the deformation parameter increases.
The complete set of thermodynamic variables, including entropy, ADM mass, and geometric volume, has been self-consistently extracted, and the validity of the first law, $d\mathcal{M} = TdS + VdP + \Phi_a da$, has been fundamentally confirmed without ad hoc modifications.

Furthermore, the modified Smarr relation, $\mathcal{M} = 2ST - 2PV + a\Phi_a$, has been successfully derived by applying Euler's theorem to the mass function.

The local thermodynamic stability has been assessed through the isobaric heat capacity, which has been shown to remain strictly positive for $a > 0$, thereby restricting the stable regime solely to the domain of positive Hawking temperature ($0 < r_{+} < \frac{9}{2aP}$).

The global thermodynamic stability has been established in the grand canonical ensemble by evaluating the Gibbs free energy, and the complete absence of a Hawking-Page phase transition has been demonstrated due to the strict negativity of $G$.

The temporal profile of the Hawking radiation has been probed by deriving the dimensionless sparsity parameter, $\eta$, based on the effective geometric cross-section. It has been discovered that increasing the $\kappa$-deformation parameter enhances the sparsity parameter, thereby amplifying the discrete nature and time intervals between successively emitted wavepackets. Finally, the asymptotic transition of the radiation toward a continuous, quasi-blackbody regime has been confirmed in the limit of high thermodynamic pressure.

\textbf{Acknowledgements:} B. Eslam Panah thanks University of Mazandaran. The authors also acknowledge the use of AI tools only for language polishing and improving the clarity of the manuscript.

\end{document}